\documentclass[aps,pre, twocolumn,showpacs,floatfix]{revtex4}

\usepackage{graphicx}
\usepackage{graphics}
\usepackage{dcolumn}
\usepackage{bm}
\begin{document}
\title{Energetics of a simple microscopic heat engine}
\author{Mesfin Asfaw}
\altaffiliation{ Present address: Max Planck Institute of Colloids and 
Interfaces, 14424 Potsdam, Germany}
\email{asfaw@mpikg.mpg.de}
\affiliation{Department of physics, Addis Ababa University\\
 P.O.Box 1176, Addis Ababa, Ethiopia}

\author{Mulugeta Bekele}%
\email{mbekele@phys.aau.edu.et}
\affiliation{Department of physics, Addis Ababa University\\
 P.O.Box 1176, Addis Ababa, Ethiopia}

\date{\today}

\begin{abstract}
We model a microscopic heat engine as a particle hopping on a
one-dimensional lattice in a periodic sawtooth potential, with or
without load, assisted by the thermal kicks it gets from
alternately placed hot and cold thermal baths. We find analytic
expressions for current and rate of heat flow when the engine
operates at steady state. Three regions are identified where the
model acts either as a heat engine or as a refrigerator or as
neither of the two. At quasistatic limit both efficiency of the
engine and coefficient of performance of the refrigerator go to
that for Carnot engine and Carnot refrigerator, respectively. We
investigate  efficiency of the engine at two operating conditions
(at maximum power and at optimum value with respect to energy and
time) and compare them with those of the endoreversible and Carnot
engines.
\end{abstract}

\maketitle

 \section  {  \bf Introduction}
Even though both macroscopic as well as microscopic heat engines
work on the same  thermodynamic principles, wide-ranging studies
have been done in improving the performance of macroscopic heat
engines \cite {and}. At present, the study of microscopic heat
engines has received considerable attention \cite{and84,aast}.
This is because of the trend in miniaturization and the need to
utilize energy resources available at microscopic scales. As such,
modelling microscopic heat engines and finding how well they
perform is a primary task to be undertaken at present.

To get a first insight as to how such engines perform, it is
important to take a toy model that has the basic ingredients. In a
recent paper \cite{mes} we considered a simple model of a Brownian
heat engine and found exact analytic expressions for quantities
like current, efficiency and coefficient of performance. This, in
turn, enabled us to explore different features of the engine such
as efficiency at maximum power, optimized efficiency as well as
efficiency at quasistatic limit. The present work addresses the
same basic issues of a tiny heat engine. However, here the
particle moves on a discrete lattice by hopping as opposed to a
continuous Brownian motion in a viscous medium. Even though the
model and its corresponding dynamics is completely different from
the one previously studied, the results can be compared with those
of the previous work at least qualitatively. As such, this work
can be taken as an independent check of the results found in the
previous work on microscopic heat engine.

The paper is organized as follows: In section II, we will first
introduce our model in the absence of external load and set up the
dynamics governing it. We will then find analytic expressions for
the steady-state current and the rate of heat produced as a
function of the model parameters. In section III, we will consider
our model in the presence of an external load and find the
steady-state current and the rate of heat flows. In section IV, expressions for
the efficiency and coefficient of performance (COP) will be
determined depending on whether the engine works as a heat engine
or as a refrigerator. Regions in the parameter space where the
model works as a heat engine, as a refrigerator and as neither of
the two will be determined. We will explore how current,
efficiency and COP behave as the model parameters vary. We will
also compare the efficiency of the engine at two operating
conditions (at maximum power and at optimum value with respect to
energy and time) with those of endoreversible and Carnot engines.
Lastly, we summarize and conclude in section V.

           \section {Zero external load}

The model we take is a modified version of the one considered by
Jarzynski and Mazonka \cite{jar} in modelling Feynmann's ratchet
and pawl system. Our model is meant to capture the motion of a
particle in a ratchet potential due to the thermal kick it gets
from periodically placed hot and cold reservoirs along the path.

Consider a particle that moves by hopping on a one-dimensional
lattice, with lattice spacing d, assisted by the thermal kick it
gets from periodically placed hot and cold reservoirs along its
path. The particle is also exposed to an external discrete
sawtooth potential which has the same period as that of the
reservoirs. In one cycle, the particle walks a net displacement of
three lattice sites, either to the right or to the left. This
corresponds to the particle crossing one sawtooth of a sawtooth
potential. The potential energy at site $i$, $U_{i}$, where $i$ is
an integer that runs from ${-\infty} $ to ${+\infty}$, is given by
\begin{equation}
  U_{i}=E[i (\mbox {mod})3 - 1].
\end{equation}
Here $E$ is positive constant having a unit of energy.

\begin{figure}
\includegraphics[scale=.4]{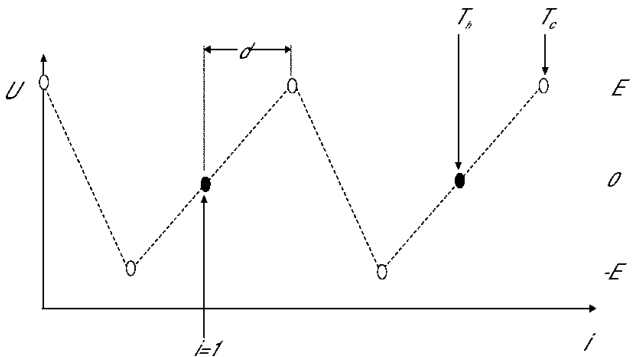}
\caption{Plot of discrete sawtooth potential without load. Sites
with dark circles are coupled to the hot reservoir ($T_h$) while
sites with open circles are coupled to the cold reservoir ($T_c$).
Site 1 is labeled explicitly and d is the lattice spacing.}
\end{figure}
The  temperature profile at site $i$, $T_{i}$, is given by
\begin{equation}
  T_{i}=\cases{
   T_{h},&if [i(\mbox {mod})3 - 1]=0,\cr
   T_{c},& \mbox{otherwise},\cr}
\end{equation}
\noindent where $T_{h}$ and $T_{c}$ are the temperatures of the
hot and cold reservoirs, respectively. Figure 1 shows the values
of sawtooth potential and of the temperature at the given lattice
sites.

 The jump of the particle from one lattice  site to  next lattice  
site
is assumed to be random in nature. The jump probability  is
determined by the amount  of energy it crosses and the temperature
of the heat reservoir to which it is coupled.  Accordingly, the
jump probability per unit time of the particle making a jump from
site $i$ to site  $i+1$ is given by $\Gamma e^{-\Delta E\over
T_{i}},$  where $\Delta E= U_{i+1} -U_{i}$ and $\Gamma $ is the
probability that the particle will attempt a jump per unit time.
We take Boltzmann's constant, $k_{B}$, to be unity. When
the particle attempts to jump, first it  decides which way to jump
(either to the left or right) with equal probability and then
jumps according to the Metropolis algorithm \cite {met}: if the
value of $\Delta E \le 0$, then the jump  definitely takes place;
if $\Delta E > 0$ then the jump takes place with probability
$exp({-\Delta E\over T_{i}})$.

The dynamics of the  model can be studied by  mapping the model to
a spin-1 particle system which exhibits identical behavior \cite
{jar}. Denoting the spin states, $s$, by $(0,\pm 1) $, the energy
function of the spin-1 particle is defined by
       \begin{equation}
        E(s)=E s.
        \end{equation}
The change of state of spin  $s$ corresponds to the jump of the
particle. If s changes from -1 to 0 or from 0 to 1 or from 1 to
-1, then this is the same as the particle jumping to the right.
The reverse process corresponds to jump to the left. The three
possible states and their corresponding energy values is shown
in the Table I.\\

\begin{table}
\begin{center}

\begin{tabular}{|l|l|l|} \hline
\hline

                 State(n)&              s&               E(s)  \\ 
\hline

                 1&                    $ -1$&             $ -E $ \\ 
\hline

                $ 2$&                    $~ 0$&                $ ~ 0  $  
\\ \hline

                 3 &                     $~ 1$ &               $~ E $ 
\\ \hline
             \end{tabular}

\end{center}
\caption{The three states of the system and their corresponding
energies}
    \end{table}

The dynamics of the particle is then described by stochastic jumps
among the three states. The process is Markovian  and we can
describe the evolution of the states with  rate equations. Let the
probability for the system to be found in state $n$ at time $t$ be
given by $p_{n}(t)$. The rate equations governing the evolution of
the three states are
\begin{equation}
 {dp_{n}\over dt}= \sum_{n'\not=n}( P_{nn'}p_{n'}-P_{n'n}p_{n}), n,
 n'=1,2,3.
\end{equation}
\noindent $ P_{n'n}$ is the transition probability rate at which
the system, originally in  state  n, makes transition to state
$n'$. Here $P_{n'n}$ is  given by the Metropolis rule. For
example,

 \begin{equation}
  P_{21}={\Gamma\over 2}e^{-E \over T_{c}},~
P_{12}={\Gamma\over 2},~
P_{32}={\Gamma\over 2}e^{-E \over T_{h}},~ P_{23}={\Gamma\over2}.
\end{equation}
\noindent In the above expressions, the factor ${1\over 2}$ is due
to the decision for the particle to jump either to the left or to
the right. The rate equation for the model can then be expressed
as a matrix equation
\begin{equation}
{d{\vec p}\over dt}=\Gamma {\bf R}{\vec p}
\end{equation}
\noindent where ${\vec p}=(p_{1},p_{2},p_{3})^{T}$. Here,  ${\bf
R}$ is a 3 by 3 matrix which is given by

\begin{eqnarray}
{\bf R}=\left(\begin{array}{ccc}               {-\mu-\mu^{2}\over 2}  &    
{1\over 2}  &   { 1\over 2} \\
               {\mu\over 2}       &     {-1-\nu\over 2} &    { 1\over2} 
\\
              {\mu^{2}\over 2}  &        {\nu\over 2}  &        -1
\end{array}
\right),
\end{eqnarray}
where $ \mu=e^{-E\over T_{c}}$ and  $\nu=e^{-E \over T_{h}}$. Note
that the sum of each column of the  matrix {\bf R} is zero,
$\sum_{m}{\bf R}_{mn}=0 $. This shows that the total probability
is conserved: ${d\over dt} \sum_
 {n} p_{n}={d\over dt}({\bf 1}^{T}.{\vec p})={\bf 1}^{T}.( \Gamma {\bf 
R}{\vec
p})=0 $.

The steady state probability distribution ${\bar {\vec p}}$ of Eq.
(4) is obtained by solving ${\bf R}{\bar {\vec p}}=0$.  We find
the normalized ${\bar {\vec p}}$ to have components given by

\begin{eqnarray}
{\bar p_{1}}&=&{1\over 1+\mu +\mu ^{2}},\\
{\bar p_{2}}&=&{2 \mu +\mu ^{2} \over
(2+\nu)(1+\mu +\mu ^{2})},\\
{\bar p_{3}}&=&{(2+\nu)(\mu ^{2}+\mu)-(2 \mu +\mu ^{2}) \over
(2+ \nu)(1+\mu+\mu^2)}.
\end{eqnarray}

The presence of the hot and cold regions along the lattice leads
to unidirectional steady state current, $J$. This steady state
current, $J$, can be found as the difference between the current
towards the right, $J^{+} $, and the current towards the left,
$J^{-}$, between any two states: $ J=J^{+}-J^{-}$. Selecting
processes taking place between states 2 and 3, the current towards
the right, $J^{+}$, is given by
\begin{equation}
J^{+}=\Gamma ({R}_{32}{\bar p_{2}}),
\end{equation}
while the current towards the left, $J^{-}$, is given by
\begin{equation}
J^{-}=\Gamma ({R}_{23}{\bar p_{3}} ).
\end{equation}
After some algebra the explicit expression for the current $J$
takes a simple form
      \begin{equation}
   J= {\Gamma \mu(\nu-\mu)\over2(2+\nu)(1+\mu+\mu^{2})}.
\end{equation}
In each cycle, the particle walks a net displacement of three
lattice sites, 3d. Therefore the  drift velocity, $v$, of the
particle is
\begin{equation}
v=3dJ.
\end{equation}
\noindent Notice that the net  current is to the right as long as
$ T_{h} >  T_{c}$ and zero when $ T_{h} = T_{c}$.

\noindent Let us next find the amount of heat transfer per cycle
between the hot and cold reservoirs as the particle climbs up or
down the potential. We assume the case where there is no energy
transfer via kinetic energy due to particle recrossing of the
boundary between the hot and cold reservoirs \cite {dereny1,dereny2}. 
When the particle jumps from state 2 to state
3, it takes heat from the hot reservoir whose amount is sufficient
to climb up the potential energy difference between the states and
equal to $E$. When the particle jumps from state 3 to state 2, it
gives heat to the hot reservoir by losing its potential energy and
is equal to $E$. Thus, the net heat per unit time taken from the
hot reservoir due to climbing up or down the potential, ${\dot
Q}_h$, is given by
\begin{equation}
 {\dot Q}_{h}=E \Gamma (R_{32}
{\bar p_{2}}- R_{23}{\bar p_{3}}).
\end{equation}
\noindent After  substituting the values of ${\bar p}_{2}$ and
${\bar p}_{3}$  from Eqs. (9)and (10) in Eq. (15), we get
\begin{equation}
 {\dot Q}_{h} =E \Gamma{ \mu (\nu -\mu)\over(
2(1+\mu +\mu^2)(2+\nu))}.
\end{equation}
When the particle jumps from state 3 to state 1 and from state 2
to state 1, it gives heat to the cold reservoir. When it jumps
from state 1 to state 3 and from  state 1 to state 2, it takes
heat from the cold reservoir. The net heat per unit time given to
the cold reservoir due to climbing up or down the potential is
given by
\begin{equation}
{\dot Q}_{c}=E \Gamma (2R_{13}{\bar p_3} -2R_{31}{\bar p_1} +
R_{12}{\bar p_2} - R_{21}{\bar p_1}).
\end{equation}
\noindent After substituting the values of ${\bar p_i}$'s, we
obtain
\begin{equation}
 {\dot Q}_{c} =E \Gamma{ \mu (\nu -\mu)\over(
2(1+\mu +\mu^2)(2+\nu))}.
\end{equation}
The above results show that ${\dot Q}_{c}={\dot Q}_{h}$. This
clearly shows that a heat taken from a hot reservoir directly goes
to the cold reservoir without doing any work. Unlike this work, in
our previous work \cite {mes}, there is work done against viscous
medium even in the absence of external load.

\noindent Let us now compute the rate of entropy production. The
rate of entropy production related with the flow of heat from the
hot reservoir  is given by ${\dot S}_{h}={{-\dot Q}_{h}\over
T_{h}}$ while the rate of entropy production related with the flow
of heat to the cold reservoir is: ${\dot S}_{c}={{\dot Q}_{c}\over
T_{c}}$. The net rate of entropy production given by ${\dot
S}={\dot S}_{h}+{\dot S}_{c}$  takes of the form,
  \begin{equation}
{\dot S}
                  ={1\over 2}\ln\left({\nu \over \mu}\right)
                   {\Gamma \mu (\nu -\mu)\over (1+\mu + 
\mu^{2})(2+\nu)}\ge
0.
\end{equation}
Notice that ${\dot S}$ is non-negative which implies that the
system is consistent with the second law of thermodynamics.

\section { Non-zero external load}

Let us consider the  model in the presence of  constant external load, 
$f$,
added to the sawtooth potential as shown in Fig. 2. Accordingly, the 
potential
energy will now be changed from $U_{i}$ to
$U_{i}+ifd$.
\begin{figure}
\includegraphics[scale=.4]{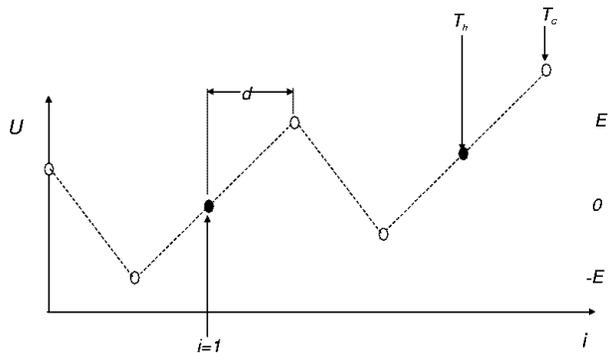}
\caption{Plot of discrete sawtooth potential with load along with
the discrete temperature profile on the lattice}
\end{figure}
The method of solving for steady state behavior is the same as
that for the load-free case.
   We consider our model when the load, $f$, is greater than zero and 
for Metropolis
   algorithm to  hold, we need to limit the range of $f$ to $0<f<{2 E 
\over d}.$
    For this range of the load, we  find  the matrix ${\bf R}$,

              \begin{eqnarray}
{\bf R}=\left (
               \begin{array}{ccc}
               -{\mu a\over 2}-{\mu^{2}\over 2a} &    {1\over 2 }&   { 
1\over 2}\\
               {\mu a\over 2}      &    {-1-\nu b\over 2}  &  { 
1\over2} \\
               {\mu^{2}\over 2a}  &      {\nu b\over 2}   &      -1
               \end{array}
                \right ),
         \end{eqnarray}

 \noindent   where
$ a=e^{-f.d \over T_{c}}$  and  $b=e^{-f.d \over T_{h}}$. Note
that the sum of each  column of the matrix ${\bf R}$ is zero,
which  shows that the total  probability is conserved. The steady
state probability  ${\bar{\vec p}}$  satisfies the matrix
equation, $ {\bf R}{\bar {\vec p}}=0. $  We solve for ${\bar{\vec
p}}$ and after normalization the final results are of the form:
\begin{eqnarray}
{\bar p_{1}}&=&{1\over 1+\mu a +{\mu ^{2}\over a}},\\
{\bar p_{2}}&=&{2\mu a +{\mu ^{2}\over a} \over
 (2+\nu b)(1+\mu a +{\mu ^{2})\over a}},\\
{\bar p_{3}}&=&{(2+\nu b)({\mu ^{2}\over a}+\mu a)-(2a \mu +{\mu
^{2}\over a}) \over(2+ \nu b)(1+\mu a+{\mu^2\over a})}.
\end{eqnarray}
\noindent Using similar approach as in section II, the expressions
for the steady state current, $ J$, and the drift velocity, $ v$,
are respectively found to be given by
\begin{equation}
 J={\Gamma \mu(ba\nu-{\mu\over a})\over 2(2+\nu b)
     (1+a\mu+{\mu^{2}\over a})},
 \end{equation}
 \noindent and
 \begin{equation}
 v= 3dJ.
\end{equation}
\noindent When the Brownian particle walks along the potential
with additional load, it takes heat from the hot reservoir and
gives some part of it to the cold reservoir and uses the rest for
climbing up the load. The difference  between the rate of heat
energy that the Brownian particle takes from the hot reservoir,
${\dot Q}_{h}$, and the rate of heat energy that it gives to the
cold reservoir, ${\dot Q}_{c}$, is the rate of useful work, ${\dot
W}$, that the particle uses to lift the load; i.e,

\begin{equation}{\dot W}={\dot Q}_{h}-{\dot
Q}_{c}=fv.
\end{equation}
\noindent From Eqs. (25) and (26), we get
 \begin{equation}
{\dot W}=fv=3fd {\Gamma \mu(ba\nu-{\mu\over a})\over 2(2+\nu
b)(1+a\mu+{\mu^{2}\over a})}.
\end{equation}
 The rate of heat energy taken from the hot reservoir by the particle, 
${\dot
 Q}_{h}$, can be obtained by a similar approach as in section II  and 
we
find its expression to be
  \begin{equation}
{\dot Q}_{h}=(E +fd){\Gamma \mu(ba\nu-{\mu\over a})\over 2(2+\nu
b)
 (1+a\mu+{\mu^{2}\over a})}.
\end{equation}
The rate of heat energy given to the cold reservoir ${\dot Q}_{c}$
is given by
 \begin{equation}
{\dot Q}_{c}=(E -2fd){\Gamma \mu(ba\nu-{\mu\over a})\over 2(2+\nu
b)
 (1+a\mu+{\mu^{2}\over a})}.
\end{equation}
The  difference between ${\dot Q}_{h}-{\dot Q}_{c}$ will be
 \begin{equation}
{\dot Q}_{h}-{\dot Q}_{c}=3fd {\Gamma \mu(ba\nu-{\mu\over a})
 \over 2(2+\nu b) (1+a\mu+{\mu^{2}\over a})},
\end{equation}
 which is exactly
equal
   to ${\dot W}$ as  given in Eq. (27).

\section  {  \bf The model as a heat engine, as a 
          refrigerator and as neither of the two}

To specify the model, one requires to specify the six quantities: 
         $ \Gamma$, $d$, $E$, $T_c$, $T_h$ and $f$. Taking $\Gamma$, 
$d$ and $T_c$ fixed we still 
          have three parameters $E$, $T_h$ and $ f$ that can be varied 
independently. We 
          convert these into three dimenssionless parameters 
$\epsilon$, $\tau$ and 
         $ \lambda$ where $\epsilon={E\over T_c}$, $\tau={T_h \over 
T_c} - 1$ and $\lambda={fd \over T_c}$. In 
          addition, we introduce a dimenssionless current $j={J\over 
\Gamma}$. We take  $T_{h}>T_{c}$ for the rest of our work.

          The current j is now a function of $\epsilon$, $\tau$ and 
$\lambda$. Figure 3 
          shows a plot of $j$ versus $\lambda$ for fixed values of 
$\epsilon$ and 
          $\tau$.
The figure shows
that the current is positive as long as the load is less than a
certain value of $\lambda$ . This corresponds to the region where
the model works as a heat engine. Using Eqs. (28), (29) and (30) the efficiency
of the heat engine, $\eta$, takes the expression
    \begin{equation}
\eta ={{\dot Q}_{h}-{\dot Q}_{c}\over {\dot Q}_{h}}= {3\lambda \over(\epsilon + \lambda)}.
\end{equation}
\noindent
On the other hand, when the load is large enough the current becomes negative
and that implies that the model works as a refrigerator. The COP, $P_{ref}$, of the refrigerator
then takes the expression
\begin{equation}
P_{ref}={{\dot Q}_{c}\over {\dot Q}_{h}-{\dot Q}_{c}}={(\epsilon-2\lambda)\over3\lambda}.
\end{equation}
\noindent
From this equation, Eq. (32), we note that in order for the model to function
as a refrigerator the upper limit for $\lambda$ must be ${\epsilon\over2}$.

\begin{figure}
\includegraphics[scale=.5]{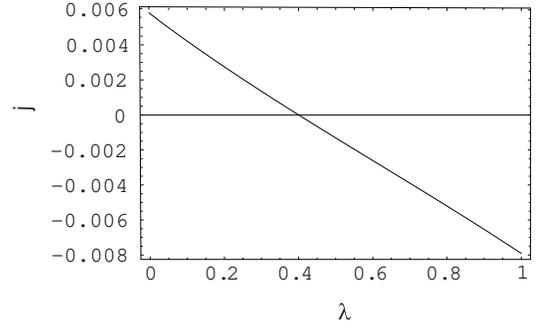}
\caption{ Plot of $j$ versus $\lambda$ for $\tau =1$ and $\epsilon=2$}
\end{figure}

The set of points in the parameter space at which current changes its direction differentiates 
the domain of operation of the model as a refrigerator from that as a heat engine.
Using the expression for $J$, Eq. (24), the value of $\lambda$ at which the current reversal
takes place is given by
\begin{equation}
\lambda={\epsilon\tau\over
 (2\tau+3) }.
\end{equation}
Note that this value of the load where current is zero is usually
called the stall force for molecular engines \cite{how}. When we
evaluate the values of both $\eta$ and $P_{ref}$ as we approach
this boundary determined by  Eq. (33), we find that they are
exactly equal to the respective values for the Carnot efficiency
and the Carnot COP: $ \displaystyle \lim _{{J\to
0^{+}}}{\eta }={(T_{h}-T_{c})\over   T_{h}} $ and
 $
\displaystyle \lim _{{J\to 0^-}}{P_{ref}}={T_{c}\over
(T_{h}-T_{c})  }$. This clearly demonstrates  that the boundary at
which current is zero corresponds to the quasistatic limit be it
from the heat engine side or from the refrigerator side.

\noindent
Let us introduce  dimenssionless parameters ${
q}_{h}={{\dot Q}_{h}\over \Gamma T_{c}}$ and  ${ q}_{c}={{\dot
Q}_{c}\over \Gamma T_{c}}$.
In analyizing the operation of our model above, we have identified that when
\begin{equation} 
0<\lambda<{\epsilon\tau\over (2\tau+3) }, 
\end{equation}
the model works as a heat engine
while it works as a refrigerator when 
\begin{equation}
{\epsilon\tau\over (2\tau+3) }<\lambda<{\epsilon\over2}.
\end{equation}
\noindent Therefore, the model neither works as a heat engine nor as a refrigerator when
\begin{equation}
\lambda>{\epsilon\over2}. 
\end{equation}
\noindent 
Figure 4a shows the three regions in the $\lambda$-$\tau$ parameters space in which 
the model operates as a heat engine, a refrigerator and as neither of the two 
for fixed value of $\epsilon=4$. On the other hand, Fig. 4b shows
these three regions in the $\lambda$-$\epsilon$ parameters space for fixed value of $\tau=1$.

\begin{figure}
  \includegraphics[scale=.42]{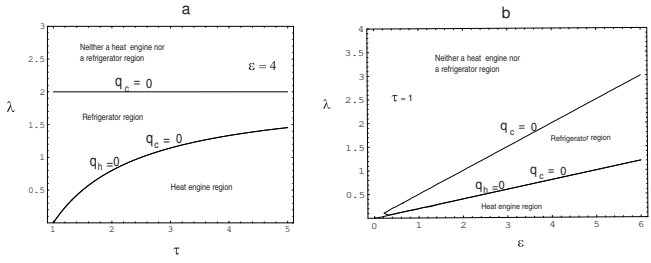}
\caption{Plots showing the three regions of operation of the model
in the (a) $\lambda$-$\tau$ paramters space for $\epsilon=4$ and
in the (b) $\lambda$-$\epsilon$ paramters space for $\tau=1$}
\end{figure}

\noindent Let us now  further investigate how the current,
efficiency and performance of the refrigerator behave as a
function of the different parameters characterizing the model.
The plot of $\eta$ versus $\lambda$
shows that,  the efficiency , $\eta$, increases with increase
in $\lambda$  until it attains its maximum value (Carnot
efficiency) as shown in Fig. 5a. On other hand, the plot of
$P_{ref}$ versus $\lambda$ shows that  within the range where the
model works as a refrigerator, coefficient of  performance of the
refrigerator, $P_{ref}$, decreases from its maximum value (Carnot
refrigerator)
 as  $\lambda$ increases (see  Fig. 5b).

\begin{figure}
\includegraphics[scale=.5]{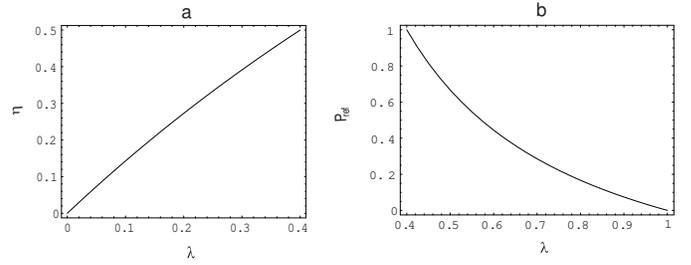}
\caption{ (a) Plot of $\eta$ versus $\lambda $ for  $\epsilon=2$  and 
$\tau =1$ (b) Plot of $P_{ref}$ versus $\lambda$ for $\epsilon=2$  and 
$\tau =1$ }
\end{figure}
\begin{figure}
\includegraphics[scale=.4]{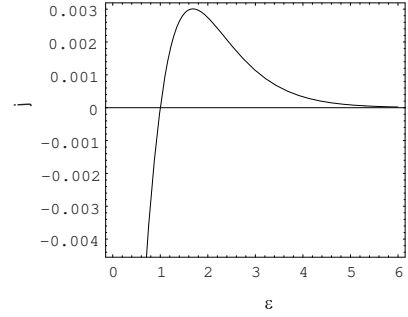}
\caption{Plot of j versus $\epsilon$ for $\lambda =.2 $ and $\tau =1$ 
}
\end{figure}

\noindent Figure 6 shows  how the current $j$ behaves as a
function of $\epsilon$. The figure shows the  presence of maximum
current at a certain finite value of $\epsilon$  for a fixed
$\tau$ and $\lambda$. This point corresponds to maximum power
delivery at which the engine operates with maximum power
efficiency, $\eta _{MP}$. Let us compare the value $\eta _{MP}$ of
our engine with the corresponding value of an endoreversible
engine. For indoreversible
engine which exchanges heat linearly at a finite rate with two
reservoirs, Curzon and Ahlborn \cite {cur,sal} showed that the
efficiency at maximum power, $\eta _{CA}$, is given by
\begin{equation} 
\eta _{CA}=1-\sqrt{\left({T_{c}\over T_{h}}\right)}.
\end{equation}
\noindent Figure 7 shows  how the
efficiency of our model heat engine when it operates with maximum
power, $\eta _{MP}$, and that of $\eta _{CA}$  behave   as a function of
$\tau$. The plots show that as $\tau$ increases the gap between
$\eta _{MP}$ and $\eta _{CA}$ decrease  and coincide at a finite
value of $\tau$. Then their difference get larger as $\tau$
increases. When we compare the two efficiencies, $\eta _{MP}$ and
$\eta _{CA}$, $\eta _{CA}$ is found by assuming linear heat
conductivity while $\eta _{MP}$ is obtained numerically without
taking any assumption. This illustrates that  $\eta _{MP}$ works
for the entire range in the allowed parameter space while   $\eta
_{CA}$ is specific and has limited significance.
\begin{figure}
\includegraphics[scale=.4]{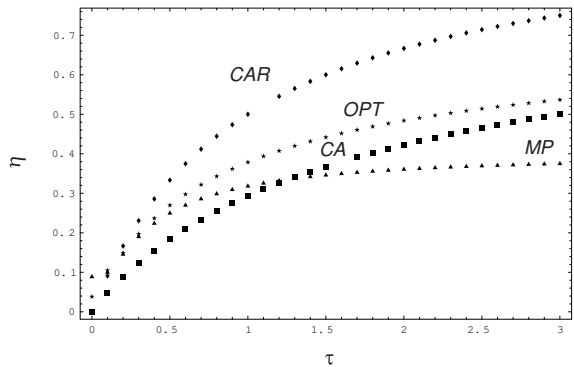}
\caption{\label{fig:epsart} Plots of ${\eta}_{CA}$,
${\eta}_{MP}$, ${\eta}_{OPT}$ and ${\eta}_{CAR}$ versus $\tau$, where 
the model
engine is put to function at $\lambda=0.2$ while $\epsilon$ is fixed 
depending on
whether it is working at either maximum power or optimized efficiency 
}
\end{figure}

Let us next compare Carnot efficiency, $\eta _{CAR}$, with that of
optimized efficiency, $\eta _{OPT}$. The  $\eta _{OPT}$   for our
model can be found using the argument stated by  Her\` nandez.
$et$ $al$ \cite{her}. We briefly  summarized the method
Her\`nandez. $et$ $al$ \cite{her} in our earlier work \cite {mes}.
The  optimized efficiency, $\eta _{OPT}$,  lies between maximum
efficiency and  efficiency under maximum power and it is given
  by optimizing
\begin{equation}
{\dot \Omega}=2{\dot W}-{(T_{h}-T_{c})\over
T_{h}}{\dot Q}_{h}.
\end{equation}
The plot of ${\dot \Omega}$ versus $\epsilon$ in Fig. 8 shows that 
the function definitely  has optimum value at finite
$\epsilon$. Evaluating the efficiency at this particular point in
the parameter space gives us the  optimized efficiency,  $\eta
_{OPT}$. We plot  $\eta _{OPT}$ and  $\eta _{CAR}$ in the same
Figure 7 that we plot the other efficiencies. The plots of $\eta
_{OPT}$ and  $\eta _{CAR}$  versus  $\tau$ show  that $\eta
_{OPT}$ lies between  $\eta _{MP}$ and  $\eta _{CAR}$. This
undeniably illustrates that the operation of the engine at
optimized efficiency is a compromise between fast transport and
energy cost.

\begin{figure}
\includegraphics[scale=.4]{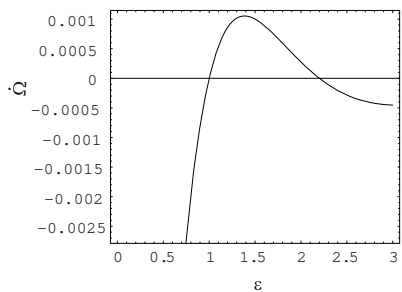}
\caption{ Plot of ${\dot\Omega}$ versus $\epsilon$ for $\lambda =.2 $ 
and $\tau =1$ }
\end{figure}

 \section  {  \bf   summary and conclusion}
In this work,  we introduced an exactly solvable model of a heat engine with
minimum ingredients. We obtained  closed form expressions for
current, efficiency and COP  of the
model. We showed that at
quasistatic limit the values of both efficiency and COP
go to that of Carnot efficiency and Carnot COP,
respectively. We then explored the basic properties
of the microscopic heat engine by varying the parameters describing the model. We further
studied the efficiency when the engine operates with maximum power
and compared this efficiency with those values one gets by
using the finite-rate linear heat exchange assumption of Curzon
and Ahlborn \cite {cur,sal}. The results of optimized efficiencies
of the  model are also reported. It is worth
to note that the particle walks  in non-viscous medium. Hence the
model does not work as a heat engine in the absence of external
load.  In the presence of external load, even though the model and
its corresponding dynamics is completely different from the one we
studied earlier, there is a qualitative agreement between this
work and the earlier work. Therefore, one can take this work as
independent check of the results we found in the previous work on
Brownian heat engine \cite {mes}.

\begin{acknowledgments}
We would like to thank The Intentional Program in Physical
Science, Uppsala University, Uppsala, Sweden for the facilities
they have provided for our research group. MA would also like to
thank Reinhard Lipowsky and Thomas Weikl for creating a pleasant
research atmosphere and enabling him finalize this work.
\end{acknowledgments}

\end{document}